\documentclass{aa}  

\usepackage{subfig}

\usepackage[labelfont=bf]{caption}

\usepackage{soul,xcolor}
\usepackage{graphicx}
\usepackage{txfonts}
\begin{document} 

\setstcolor{red}

   \title{Statistics of solar wind electron breakpoint energies using machine learning techniques}

   \author{M. R. Bakrania
          \inst{1},
          I. J. Rae
          \inst{1},
          A. P. Walsh
          \inst{2},
          D. Verscharen
          \inst{1,3},
          A. W. Smith
          \inst{1},
          T. Bloch
          \inst{4}
          \and
          C. E. J. Watt
          \inst{4}
          }
   \authorrunning{M. R. Bakrania et al.}

   \institute{Department of Space and Climate Physics, Mullard Space Science Laboratory, University College London, Dorking, RH5 6NT, UK\\
              \email{mayur.bakrania.14@ucl.ac.uk}
         \and
             European Space Astronomy Centre, Urb. Villafranca del Castillo, E-28692 Villanueva de la Cañada, Madrid, Spain
         \and
             Space Science Center, University of New Hampshire, Durham, NH 03824, USA
         \and
             Department of Meteorology, University of Reading, Reading, RG6 6AE, UK
             }


 
  \abstract{Solar wind electron velocity distributions at 1 au consist of a thermal `core' population and two suprathermal populations: ‘halo’ and ‘strahl’. The core and halo are quasi-isotropic, whereas the strahl typically travels radially outwards along the parallel or anti-parallel direction with respect to the interplanetary magnetic field. Using Cluster-PEACE data, we analyse energy and pitch angle distributions and use machine learning techniques to provide robust classifications of these solar wind populations. Initially, we used unsupervised algorithms to classify halo and strahl differential energy flux distributions to allow us to calculate relative number densities, which are of the same order as previous results. Subsequently, we applied unsupervised algorithms to phase space density distributions over ten years to study the variation of halo and strahl breakpoint energies with solar wind parameters. In our statistical study, we find both halo and strahl suprathermal breakpoint energies display a significant increase with core temperature, with the halo exhibiting a more positive correlation than the strahl. We conclude low energy strahl electrons are scattering into the core at perpendicular pitch angles. This increases the number of Coulomb collisions and extends the perpendicular core population to higher energies, resulting in a larger difference between halo and strahl breakpoint energies at higher core temperatures. Statistically, the locations of both suprathermal breakpoint energies decrease with increasing solar wind speed. In the case of halo breakpoint energy, we observe two distinct profiles above and below 500 km/s. We relate this to the difference in origin of fast and slow solar wind.}

   \keywords{plasmas --
                methods: statistical --
                Sun: solar wind
               }

   \maketitle
%

\section{Introduction}

Solar wind electron velocity distributions at 1 au consist of three main populations: the thermal ($<$50 eV) population, termed the core, and two suprathermal ($\sim$60–1000 eV) populations termed the halo and the strahl \citep{1975,maksimovic}. The core has an average temperature at 1 au of $\sim$10\textsuperscript{5} K \citep{balogh} and exhibits a nearly Maxwellian velocity distribution. At 1 au, the core contains $\sim$95\%-96\% of the total solar wind electron density in slow wind \citep{mccomas,maksimovic,Stverak} and $\sim$90\% in fast wind \citep{Stverak}. The halo, on the other hand, exhibits a $\kappa$-distribution and forms tails in the total electron velocity distribution. The $\kappa$-distribution has a similar shape to the Maxwellian distribution at low thermal velocities. At speeds greater than the thermal speed, the $\kappa$-distribution decreases as a power law. The $\kappa$-distribution of the halo has a greater temperature than the Maxwellian distribution of the core \citep{1975}. The core and halo are quasi-isotropic populations, whereas the strahl travels along the interplanetary magnetic field (IMF) and can be observed in either the parallel or anti-parallel magnetic field direction \citep{1978}, or in both directions \citep{1987,owens_2017}, depending on the IMF topology. There are also times in which a strahl population is not detectable \citep{anderson}, particularly in slow solar wind \citep{Gurgiolo}.

The thermal core is thought to form in the corona, as a result of Coulomb collisions and wave-particle interactions \citep{Pierrard2001,vocks}. Likewise, suprathermal solar wind electrons originate from the solar corona \citep{Vinas_2000,Che_2014} and then evolve into the strahl and halo populations as they travel away from the Sun. The majority of the halo population is formed by the scattering of strahl electrons via Coulomb collisions \citep{horaites} and wave-particle interactions \citep{gary_94,Landi_2012,Vasko_2019,Tong_2019,Verscharen_2019} as it travels outwards in the solar wind \citep{saito_gary,pagel}. The strong field-aligned nature of the strahl occurs due to adiabatic focusing effects \citep{Owens2013}, which are particularly prevalent at smaller distances from the Sun due to larger gradients in magnetic field strength. Adiabatic focusing describes the change in pitch angle experienced by an electron that slowly travels into a region with a stronger or weaker magnetic field. Ignoring any scattering effects, an electron's pitch angle evolution with heliocentric distance then depends on the conservation of its magnetic moment. This conservation law results in a decrease in pitch angle with increasing heliocentric distance \citep{parker,owens}.

At 1 au, suprathermal electrons do not undergo any significant Coulomb collisions \citep{Vocks_2005}. This suggests that adiabatic focusing is the dominant mechanism experienced by these electrons. Under this assumption, the strahl narrows with heliocentric distance into a collimated beam of width $<$1$^{\circ}$ \citep{anderson}. However, the strahl has been observed to broaden to pitch angles of greater than 20$^{\circ}$ at 1 au \citep{Hammond,anderson,Graham_17}, suggesting the presence of additional scattering processes \citep{bercic}. This increase in strahl width with radial distance is not constant, as observations at both 5.5 au and 10 au show that the rate of solar wind electron pitch angle scattering decreases with radial distance \citep{walsh_strahl,Graham_17}.

The strahl and halo relative number density ratios vary with radial distance. We use $n_{s}$, $n_{h}$, $n_{c}$ and $n_{e}$ to define the strahl, halo, core and total electron number densities respectively. The ratio $(n_{s}+n_{h})/n_{e}$ stays approximately constant with heliocentric distance in both fast and slow wind, according to \cite{Stverak}, who obtains physical parameters by fitting to electron velocity distributions. The effect of strahl broadening results in a decrease of $n_{s}/n_{e}$ with increasing heliocentric distance. Concurrently, $n_{h}/n_{e}$ increases with heliocentric distance \citep{Stverak}, further indicating a link between the strahl and halo, and that the relevant scattering mechanisms cause the strahl to broaden and eventually scatter into the halo.

Multiple studies \citep[e.g.][]{1975,Scudder,Pilipp,mccomas_92,Stverak} identify the energy above which non-thermal parts of the distribution deviate from the Maxwellian core. We define this energy as the `breakpoint energy', $E_{bp}$. Particles above a certain energy experience minimal collisions, creating the non-thermal tails in the electron velocity distribution function and forming halo and strahl. This `breakpoint energy' is thought to be determined primarily by Coulomb collisions \citep{Scudder}. Based on the properties of Coulomb collisions and the inhomogeneity of the solar wind, and assuming minimal wave-particle interactions in the heliosphere, this breakpoint energy theoretically relates to core temperature, $T_{c}$, and heliocentric radial distance, $r$, as \citep{Scudder}:
\begin{equation}
E_{bp}(r) = 7k_{B}T_{c}(r).
\label{eq:bp}
\end{equation}
At 1 au, the average breakpoint energy is $\sim$ 60 eV \citep{1975}, however, its value varies with the local core temperature and solar wind speed \citep{Stverak}. The breakpoint energies between core and halo and between core and strahl are often different. Using electron velocity distribution functions, \cite{Stverak} show that the ratio between halo breakpoint energy and core temperature is larger than the ratio between strahl breakpoint energy and core temperature, across a range of heliocentric distances. At 1 au, \cite{Stverak} observe $E_{bp}/k_{B}T_{c}$ $\approx$ 6.5 and $E_{bp}/k_{B}T_{c}$ $\approx$ 4.5 for halo and strahl respectively. Empirical studies based on Ulysses data at heliocentric distances $>$ 1 au \citep{mccomas_92} find that the breakpoint energy decreases with distance $\propto r^{-0.4}$, and ranges between 47 eV and 60 eV at 1 au, and that $E_{bp}/k_{B}T_{c}$ $\approx$ 7.5. However, due to the differences in the applied methods for the determination of the cut-off between core and suprathermal distribution functions, this difference is not significant.

Models, which assume an absence of exchange between parallel and perpendicular pressure, predict a core temperature anisotropy in the slow solar wind of $T_{c\parallel}/T_{c\perp}$ $\approx$ 30, where $T_{c\parallel}$ and $T_{c\perp}$ are the temperature of the core components in the direction parallel and perpendicular to the magnetic field respectively \citep{Phillips_anisotropy}. Observations at 1 au, however, find a temperature anisotropy, $T_{c\parallel}/T_{c\perp}$ $\approx$ 1.2 \citep{1975,pilipp_anisotropy}. To explain this discrepancy between theory and observations, electron instabilities driven by temperature anisotropy, Coulomb collisions, and heat-flux skewness are thought to transfer the internal electron kinetic energy from the parallel to perpendicular direction \citep{pilipp_anisotropy}. \cite{stverak_anisotropy} shows that  $T_{c\perp}/T_{c\parallel}$ = 0.75$\pm$0.15 in fast wind streams, which is also consistent with the parallel to perpendicular transfer of internal kinetic energy.

In this paper, we demonstrate how machine learning techniques such as clustering can be applied to solar wind electron data, and we discuss its advantages over previous traditional methods, which involve fitting to electron velocity distributions. In order to demonstrate specific advantages, we analyse a particular physical property of solar wind electron populations - the breakpoint energy - by identifying core, halo, and strahl distributions at 1 au. Characterising the breakpoint energy is important as this property of a distribution function provides a diagnostic of the relative importance of scattering mechanisms such as Coulomb collisions and wave-particle interactions. These mechanisms determine the shape of electron distribution functions in both solar wind and astrophysical plasmas \citep[e.g.][]{astrophysical_plasmas,Pilipp_evdf}. In addition to these benefits, understanding the location of this cut-off between the thermal and non-thermal parts of a distribution, using only a statistical analysis of the data, provides useful limiting parameters for future studies which require multi-component fits to the total electron velocity distribution \citep{bercic}.

Machine learning provides us with a robust method of classification from which fine variations of electron populations in relation to energy and pitch angle can be derived, with the advantage of not requiring prior assumptions of the distributions of these populations. Applying machine learning techniques to a large dataset builds upon previous empirical studies of the suprathermal breakpoint energy. By classifying individual electron distributions, we characterise solar wind electron populations on a higher energy resolution than previous studies. As a result, our method enables breakpoint energy to be explored further with respect to other solar wind parameters, and by doing so we draw physical conclusions based on the relationship between this fundamental property and each parameter, for both the halo and the strahl. Machine learning techniques will become increasingly important with the anticipated volume of high cadence electron data from, for example, the Solar Orbiter mission \citep{solar_orbiter}.

\section{Method}

In this section, we describe the steps we take in order to classify solar wind electrons with machine learning techniques, followed by a description of the validation of our method. Firstly, we (1) determine which spacecraft and instruments are best suited for this study, and locate data from different solar wind regimes for testing. Secondly, we (2) identify possible machine learning models to be used to distinguish between electron populations. We then (3) verify the use of these models to find the `breakpoint energy' between suprathermal and core electrons. Following on, we (4) apply these machine learning algorithms to separate halo and strahl electrons based on their energy and pitch angle distributions. Lastly, we (5) calculate relative number densities of each population for different solar wind speeds and compare to previous studies \citep{Stverak}. This allows us to determine the effectiveness of our machine learning models.

Steps 3 and 4 are particularly important for our statistical study. We use the method in step 3 to calculate the breakpoint energy in each pitch angle bin and then step 4 to predict whether the strahl or halo is dominant at that pitch angle.

\subsection{Data}

We used data \citep{csa} from the PEACE \citep[Plasma Electron And Current Experiment,][]{peace,faz_peace} instrument onboard the Cluster mission's C2 spacecraft \citep{escoubet_2001}. Cluster consists of four spacecraft, in tetrahedral formation, each spinning at a rate of 4 s\textsuperscript{-1}. The PEACE data are recorded with a 4 s time resolution and are based on two instantaneous measurements of the pitch angle distribution per spin. The dataset is a two-dimensional product containing twelve 15$^{\circ}$ wide pitch angle bins and 44 energy bins, spaced linearly between 0.6 eV to 9.5 eV and logarithmically at higher energies. PEACE works by simultaneously recording elevation bins at two specific azimuth angles separated by 180$^{\circ}$. We initially corrected the PEACE data for spacecraft potential by using measurements from the Cluster-EFW instrument \citep{gustafsson_efw} and corrections according to the results of \cite{cully}. We discarded data from energy bins below the calculated spacecraft potential. 

We used the solar wind speed measurements from the Cluster-CIS instrument onboard the C4 spacecraft \citep{reme_2001}, while the position and magnetic field measurements are taken from the Cluster-FGM instrument \citep{cluster_fgm}. Using the CIS measurements, we initially separated our input electron pitch angle distribution data into three (fast, medium and slow) solar wind regimes to test our machine learning models. These regimes cover roughly 1-2 hours of data and have average solar wind velocities of 686 km/s, 442 km/s and 308 km/s. The time periods we identify with these fast, medium and slow wind regimes are 08:51-10:19 (02/03/2004), 00:38-01:35 (30/01/2003) and 04:33-06:18 (08/02/2009), respectively \citep{kajdic}. We use these specific time intervals since they contain enough data points ($>$ 10,000 samples) to effectively train and test our machine learning models.

\subsection{Machine Learning Techniques}

We predominantly used unsupervised learning algorithms to determine breakpoint energies, as well as separate halo and strahl. Unsupervised learning algorithms do not require `training' so they are more time efficient than supervised learning algorithms. Our choice of algorithm is the K-means clustering method \citep{k-means} from the scikit-learn library \citep{scikit-learn}. Unsupervised learning algorithms have the advantage of not needing the user to assign labels to training data, which reduces bias and allows large surveys to be carried out more efficiently. In the K-means algorithm, the number of clusters, K, is manually set to 2 to reflect the number of populations we aim to distinguish between: a core cluster and a suprathermal cluster. To calculate the breakpoint energy at a specific pitch angle, our algorithm sorts between energy distributions, at that pitch angle, and separates the distributions into two groups on either side of the determined breakpoint energy. We define $x_{i}$ as the vector representation of the phase space density (PSD) tuples, where the index $i$ labels tuples of three subsequent energy bins (i.e. energy distributions spanning three energy bins). We define $\mu_{j}$ as the vector representation of two random PSD tuples, where the index $j$ labels each cluster. The algorithm sorts these energy distributions into clusters by minimising the function:
\begin{equation}
\sum_{i=1}^{n}\sum_{j=1}^{K=2}\omega _{ij}\left \| x_{i}-\mu _{j} \right \|^{2},
\label{eq:k-means}
\end{equation}
where
\begin{equation}
\mu_{j} = \frac{\sum_{i=1}^{n}\omega _{ij} x_{i}}{\sum_{i=1}^{n}\omega _{ij}},
\label{eq:k-means_mu}
\end{equation}
\begin{equation}
\omega_{ij} = \begin{cases}
1 & \text{ if } x_{i}\text{ belongs to cluster }j \\
0 & \text{ otherwise,} 
\end{cases}
\label{eq:k-means_weight}   
\end{equation}
and $n$ is the number of 3-tuples at a fixed pitch angle. As each 3-tuple overlaps with its neighbouring 3-tuples, $n=N_{e}-2$, where $N_{e}$ is the number of energy bins at each pitch angle. By minimising the function in Eq. \eqref{eq:k-means}, our algorithm calculates the breakpoint energy by: (1) randomly selecting two PSD vectors in the dataset to become the central points of each cluster, $\mu_{j}$, known as centroids, (2) assigning all remaining PSD vectors, $x_{i}$, to the closest centroid, based on the least-square error between each vector and the centroids, (3) computing new centroids, $\mu_{j}$, by calculating the average vector representation of the PSD vectors assigned to the previous centroid, (4) reassigning each PSD vector, $x_{i}$, to the new nearest centroid, $\mu_{j}$, and (5) iterating steps 3 and 4 until no more reassignments occur.

Once the two clusters have been finalised, the breakpoint energy at the relevant pitch angle is determined to be the midpoint between the uppermost energy bin in the cluster of 3-tuples associated with lower energies (which represents the core), and lowest energy bin in the cluster of 3-tuples associated with higher energies (which represents suprathermal electrons). As the PSD decreases with increasing energy in the relevant energy range, we are able to locate a clear boundary between the two clusters. To separate strahl and halo electrons, we use energy distributions in conjunction with pitch angle distributions, as discussed below in Section \ref{sec:Separating Halo and Strahl}. The process of applying our K-means algorithm to pitch angle distributions is analogous to the method described above, with $x_{i}$ now representing a pitch angle distribution at a certain energy, however in this case we find the `break' in pitch angle instead. A detailed account of how the K-means algorithm works is provided by \cite{k-means}.

We validate our clustering method by comparing test cases to an accurate supervised learning algorithm, trained on a subset of manually labelled (as halo or strahl) pitch angle and energy distributions. Once trained, the supervised learning algorithm predicts which class (halo or strahl) a new pitch angle or energy distribution belongs to. We compare supervised learning algorithms by calculating their ROC (Receiver operating characteristic) scores \citep[e.g.][]{precision}. The ROC score compares a binary classification model's sensitivity (true positive rate) and specificity (1 - false positive rate) performance. We find the K-Nearest Neighbours (KNN) \citep[e.g.][]{k-nearestneighbour} algorithm performs best, achieving ROC scores $>$ 90\% in all tests. This model classifies data by finding the `majority vote' of the nearest (labelled) neighbours to each unclassified data-point.

\subsection{Distinguishing Between Suprathermal and Core Electron Populations}

\begin{figure}[ht]
\centering
\includegraphics[width=8cm]{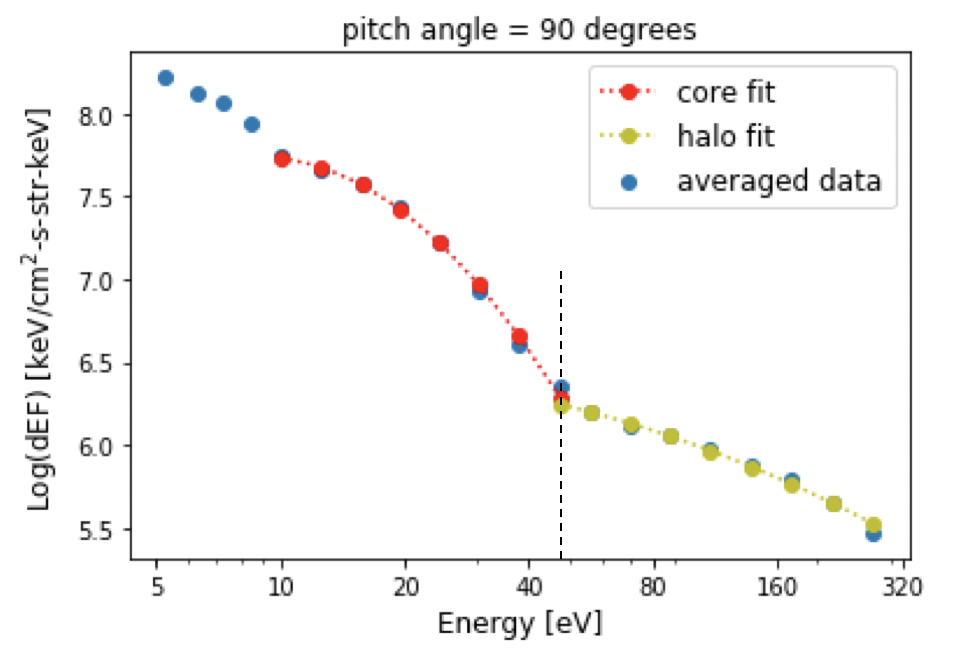}
\caption{Differential energy flux as a function of energy at 90$^{\circ}$, averaged across times 08:51-10:19 (02/03/2004) during our fast wind regime. The red curve represents a fit to the core electron energy range and the yellow curve to the halo energy range. The grey dashed line marks the so-called `breakpoint energy' at 45$\pm$3 eV.}
\label{fig:bp_90_dash}
\end{figure}

We demonstrate the use of unsupervised clustering to calculate the breakpoint energy. Figure \ref{fig:bp_90_dash}, which shows a cut of the differential energy flux distribution at constant pitch angle, visualises this breakpoint energy. Figure \ref{fig:bp_90_dash} contains three regions with different distribution functions. At energies below the spacecraft potential at $\sim$10 eV, photo-electrons dominate (blue dots). At slightly higher energies, between 10 eV and $\sim$45 eV, the distribution represents core electrons. At larger energies we observe the halo population. We fit a Maxwellian (red) and $\kappa$-distribution (yellow) \citep{Stverak} to the core and halo respectively, to determine the energy at which the distributions intersect, that is, the `breakpoint energy'. 

The intersection in Figure \ref{fig:bp_90_dash} results in an estimated halo breakpoint energy of 45$\pm$3 eV. We apply the same method to flux measured at pitch angles 0$^{\circ}$ and 180$^{\circ}$, where the strahl carries the highest value of the flux density in the suprathermal energy regime. These intersections show a separation between the core and suprathermal strahl population at 42$\pm$3 eV. We use the core-halo intersection in Figure \ref{fig:bp_90_dash}, which is labelled by the dashed line, to validate our use of clustering analysis to calculate breakpoint energy, detailed below. 

We omit energies below 10 eV and above 540 eV from our dataset and use the K-means clustering algorithm \citep{k-means} to classify the suprathermal and core populations, and hence calculate the breakpoint energy, at our choice of pitch angle. We assess the algorithm's performance by comparing its classifications of the core population at each time step to an averaged distribution of the data, such as in Figure \ref{fig:bp_90_dash}. This unsupervised learning method produces encouraging results. At 90$^{\circ}$ pitch angle, the algorithm estimates the average breakpoint energy to be 45 eV$\pm$3. The accuracy score between algorithm's classifications and a fit to the averaged distribution is 92.9$\%$. As we predict binary classifications, we consider metric scores close to 90\% as `good' scores when testing our models, based on what previous studies achieve \citep[e.g.][]{qian_accuracy,zhang_accuracy}. 

\subsection{Separating Halo and Strahl Electrons}
\label{sec:Separating Halo and Strahl}

\begin{figure*}[ht]
\centering
\includegraphics[width=11cm]{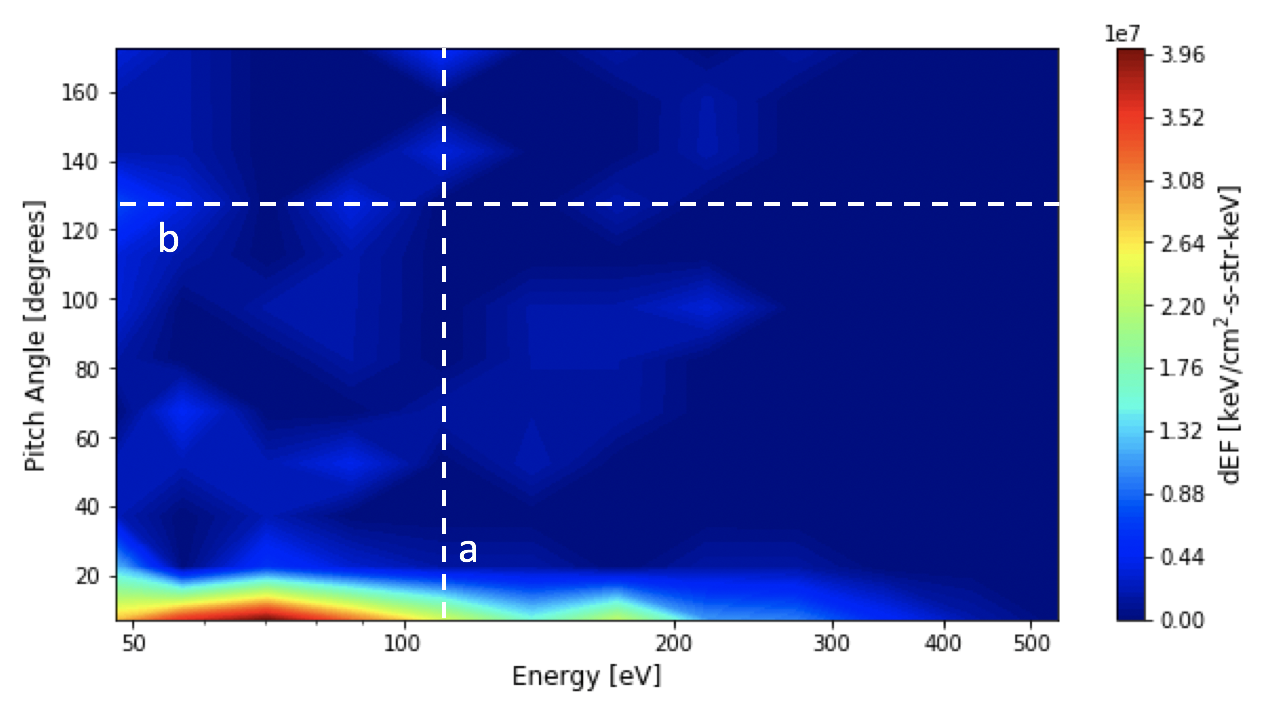}
\caption{Two-dimensional colour plot of the measured electron differential energy flux, across a 4 second window (08:57:28-08:57:32 on 02/03/2004) during our fast wind regime. The data are plotted as a function of pitch angle (degrees) and energy (eV), across an energy range of $\sim$44 eV to $\sim$540 eV. The vertical and horizontal white dashed lines represent where cuts are made to obtain: a) the pitch angle distribution at 110.09 eV, and b) the energy distribution at 127.5$^{\circ}$.}
\label{fig:2d_strahl_t500_labels}
\end{figure*}

Figure \ref{fig:2d_strahl_t500_labels} illustrates a typical differential energy flux distribution as a function of pitch angle and energy distribution for one particular time (08:57:28-08:57:32 on 02/03/2004) recorded by Cluster-PEACE. We limit the energy range to the suprathermal energy regime, as a result of our breakpoint energy analysis.

\begin{figure*}[ht]
    \centering
    \subfloat[Pitch Angle Distribution at 110.09 eV]{{\includegraphics[width=6.5cm]{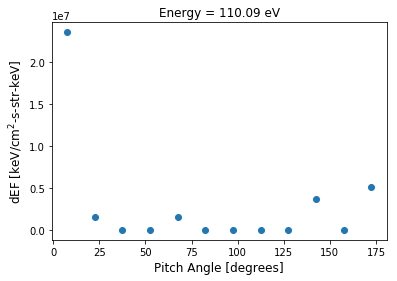}}
    \label{fig:strahl_pad_t500}}
    \qquad
    \subfloat[Energy Distribution at 127.5$^{\circ}$]{{\includegraphics[width=6.5cm]{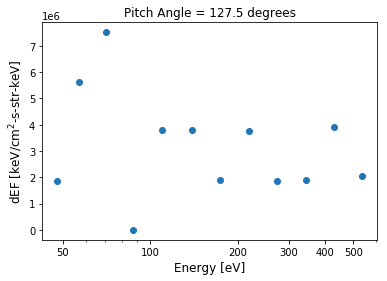}}
    \label{fig:halo_fed_nolog}}%
    \caption{a) Pitch angle distribution at an energy of 110.09 eV and, b) energy distribution at a pitch angle of 127.5$^{\circ}$ as projected from the vertical and horizontal white lines in Figure \ref{fig:2d_strahl_t500_labels}.}
    \label{fig:pad_and_fed}%
\end{figure*}

In order to show the average pitch angle distribution (PAD), we take vertical slices in Figure \ref{fig:2d_strahl_t500_labels} at a given energy. The white line (a) in Figure \ref{fig:2d_strahl_t500_labels} represents the slice from which we obtain the example PAD in Figure \ref{fig:strahl_pad_t500}. Below the typical breakpoint energy these distributions are relatively isotropic across all pitch angles, which is in contrast to the strahl distribution \citep{mccomas_92}. At higher energies within the suprathermal regime, PADs either show a quasi-isotropic distribution, which represents the halo, or an anisotropic distribution with peak fluxes recorded at 0$^{\circ}$ and/or 180$^{\circ}$, which represents the halo population at all pitch angles overlaid with field-aligned strahl. 

From our breakpoint energy analysis, we limit our input data to energies above 44 eV and convert these suprathermal data to PADs across our energy range, e.g. as shown in Figure \ref{fig:strahl_pad_t500}. We use an arbitrary 10-minute subset of time intervals, equivalent to 1800 samples, as training data. We assign each PAD a label, depending on whether strahl is or is not present. Subsequently, the entire set of PADs during our chosen wind speed regime are classified, based on a trained KNN model. We find a strong agreement between this supervised method and using K-means to cluster the fast wind set of PADs into two groups (halo and strahl), with a calculated ROC score of 90.3\%.

Classifying PADs informs us of whether a strahl is present at a certain energy, however we require classification of the energy distributions at each pitch angle to extract the width of the strahl. The white line (b) in Figure \ref{fig:2d_strahl_t500_labels} represents the slice from which we obtain the example energy distribution in Figure \ref{fig:halo_fed_nolog}. We now use a 10-minute interval of energy distributions, at each pitch angle, for our training data and provide labels depending on whether strahl is present or not at that pitch angle. We find a strong similarity between the supervised and unsupervised methods, when classifying the entire set of flux-energy distributions, with a ROC score of 98.3\%. This comparison therefore validates the use of the unsupervised method for any larger statistical survey.

\begin{figure*}[ht]
\centering
\includegraphics[width=11cm]{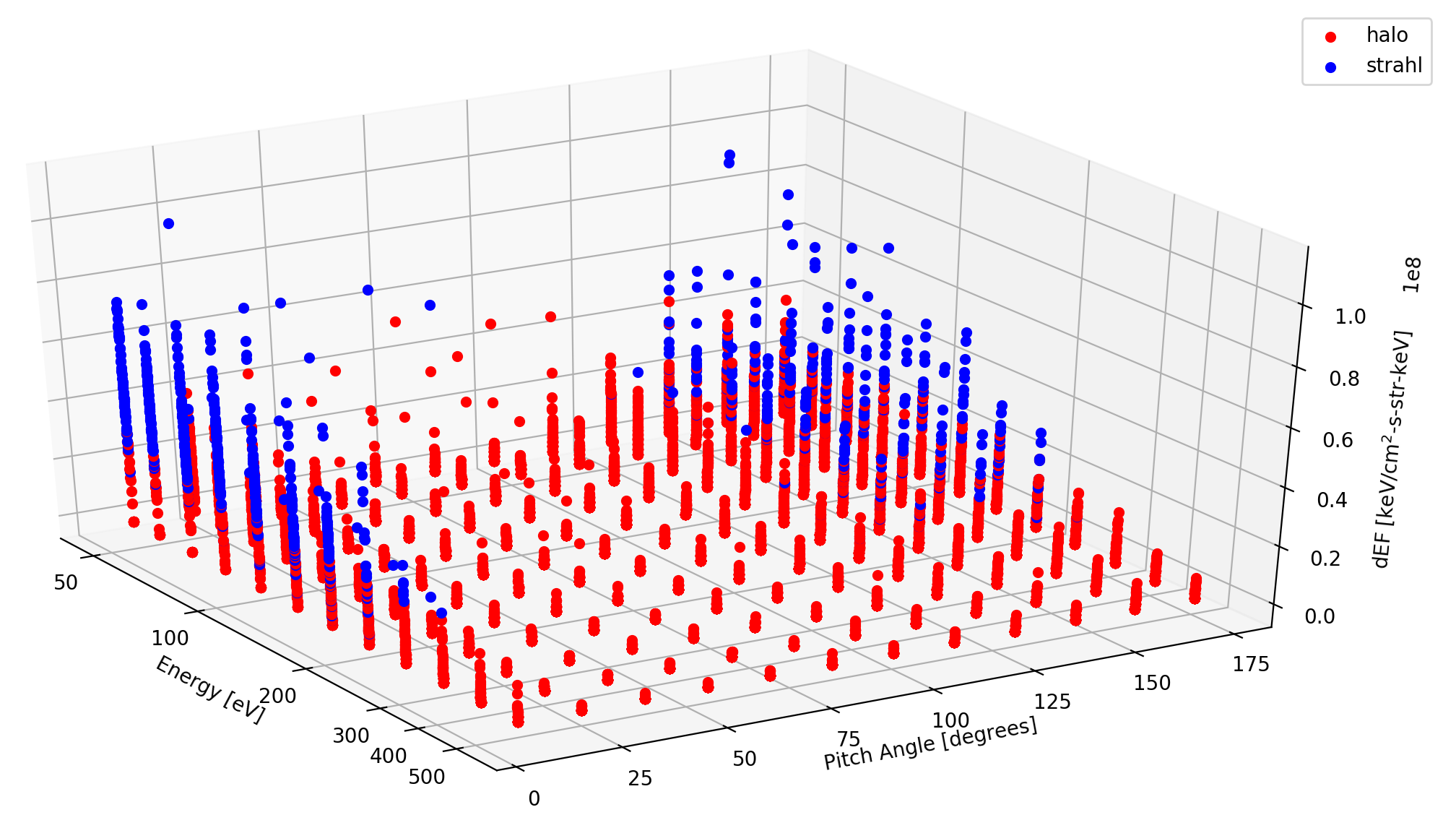}
\caption{3D scatter plot of the differential energy flux as a function of pitch angle and energy, for the fast solar wind dataset. The colours define whether the K-means clustering algorithm labels each bin as either containing strahl and halo flux (blue) or only halo flux (red).}
\label{fig:3d_classification_plot}
\end{figure*}

For each time step, we combine the classifications of suprathermal PADs and suprathermal energy distributions to create a grid detailing whether the measured flux in each energy and pitch angle bin is dominated by halo electrons or by strahl electrons. A bin is identified as containing strahl if both the PAD and energy distribution it resides in are classed as strahl by the K-means algorithm. We show the results of our strahl and halo classification in fast wind in Figure \ref{fig:3d_classification_plot}. Each point represents a single measurement at a given pitch angle and energy, with the colour depicting the class (halo or strahl). The higher fluxes near 0$^{\circ}$ and 180$^{\circ}$ are associated with strahl (blue points). On occasion, broader strahl is detected, as illustrated by the presence of blue points at higher fluxes near 75$^{\circ}$. The existence of red points across all pitch angles at lower fluxes confirms the presence of the halo as an isotropic population.

\begin{figure*}[ht]
\centering
\includegraphics[width=11cm]{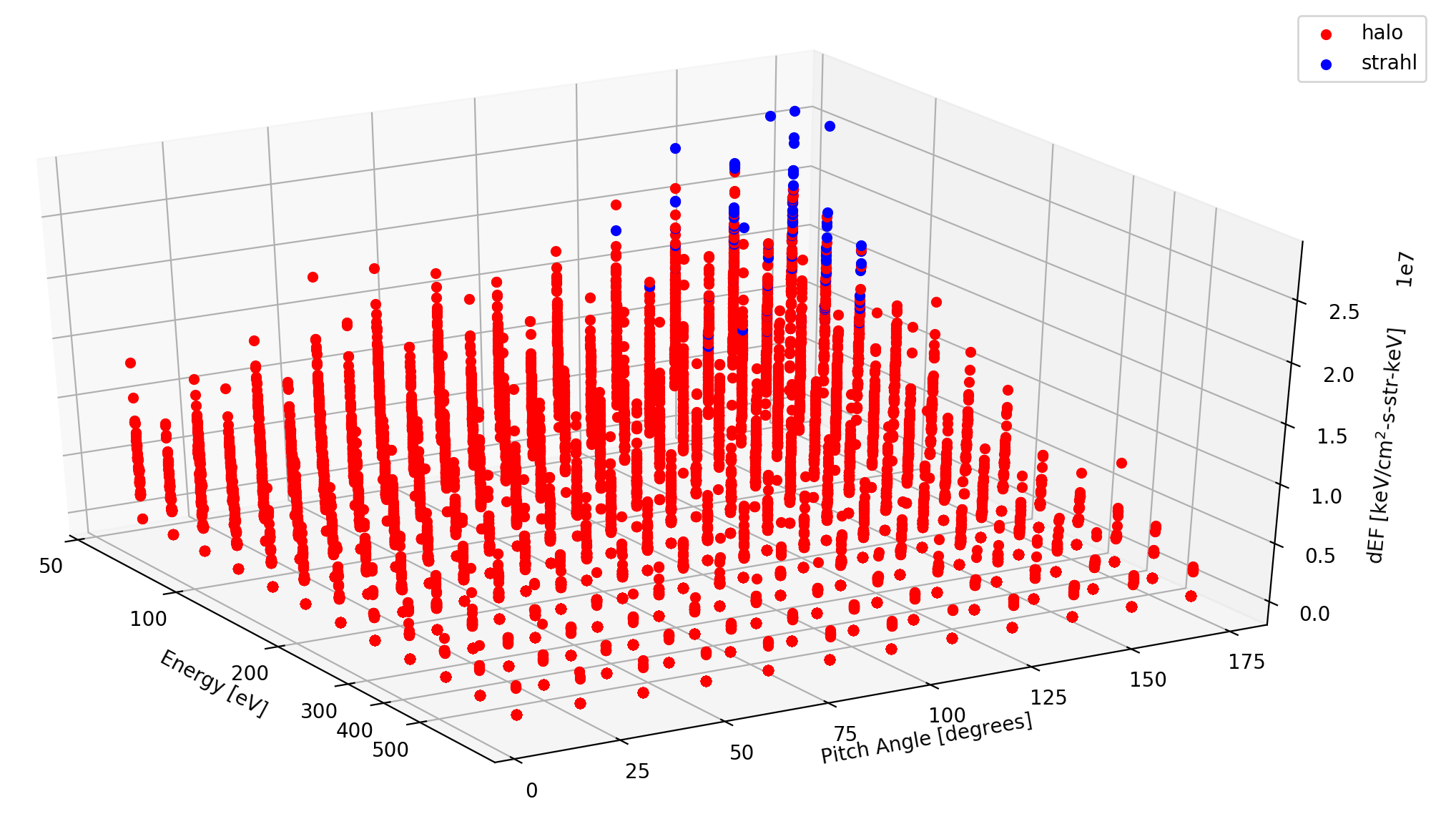}
\caption{3D scatter plot of the differential energy flux as a function of pitch angle and energy, for the slow solar wind dataset.}
\label{fig:3d_classification_slow_wind}
\end{figure*}

We show the results of our strahl and halo classification in slow wind in Figure \ref{fig:3d_classification_slow_wind}. We see that the number of blue points, associated with the strahl, is much reduced in the slow wind than in the fast wind (see Figure \ref{fig:3d_classification_plot}). This finding is consistent with the observed lower occurrence of strahl during times of slow solar wind \citep[e.g.][]{Gurgiolo}. Both Figures \ref{fig:3d_classification_plot} and \ref{fig:3d_classification_slow_wind} confirm that only halo electrons exist at pitch angles around 90$^{\circ}$. We see for both fast and slow wind cases that the strahl exhibits higher differential energy fluxes than the halo. The scattering of strahl electrons into the halo results in a larger spread of electrons across all pitch angles, decreasing the peak flux at any one pitch angle.

\subsection{Calculating Relative Number Densities}

After classifying the dataset into core, halo and strahl regions, we calculate the differential energy flux attributed to each population. In order to account for halo electrons in strahl pitch angle and energy bins, we subtract the halo flux, averaged over all pitch angles at a fixed energy, from strahl fluxes at that energy and assign it to the total halo flux. Differential energy flux relates to the partial number density (cm\textsuperscript{-3}) of each electron population as according to Eq. \eqref{eq:conversion} \citep{calibration}:
\begin{equation}
\Delta n\approx 5.4\times 10^{-10} \; E^{-\frac{3}{2}} \; \Delta E \; \Delta \Omega \; J \; \; (\textrm{cm}^{-3}),
\label{eq:conversion}
\end{equation}
where $E$ is the average energy within interval $\Delta E$ (both measured in keV/$Q$) and $J$ is the average differential energy flux (keV/cm\textsuperscript{2}-s-str-keV) at energy $E$. $\Delta \Omega$ is the solid angle ($\leq$4$\pi$) over which $J$ is measured and relates to the pitch angle widths. 

\begin{figure}[ht]
\centering
\includegraphics[width=\hsize]{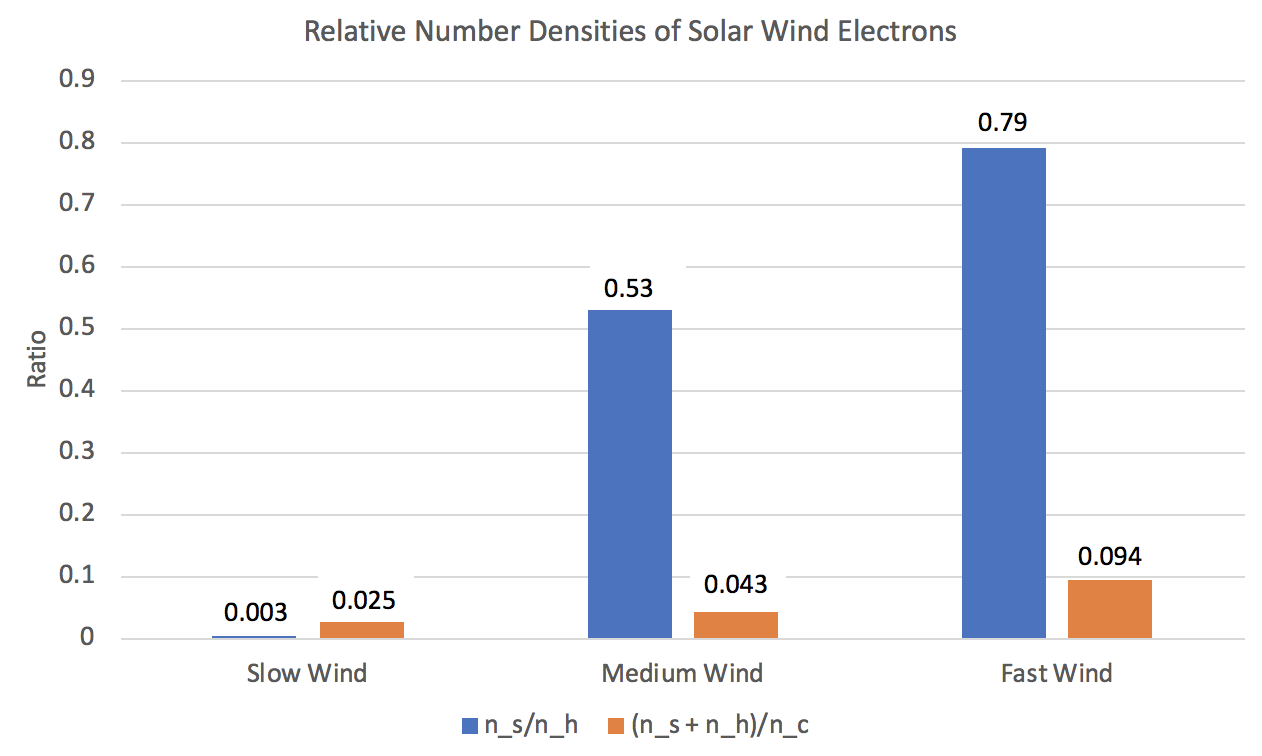}
\caption{$n_{s}/n_{h}$ and $(n_{s}+n_{h})/n_{c}$ ratios for slow, medium and fast solar wind.}
\label{fig:wind_speed_ratios_labels}
\end{figure}

In Figure \ref{fig:wind_speed_ratios_labels}, we show the conversion of differential energy flux to number density. In slow wind: the ratio $n_{s}/n_{h}$ = 0.003 and $(n_{s}+n_{h})/n_{c}$ = 0.025 where $n_{s}$, $n_{h}$ and $n_{c}$ represent the strahl, halo and core number densities. In intermediate wind: $n_{s}/n_{h}$ = 0.53 and $(n_{s}+n_{h})/n_{c}$ = 0.043 while in fast wind: $n_{s}/n_{h}$ = 0.79 and $(n_{s}+n_{h})/n_{c}$ = 0.094.

Our calculated densities are of the same order as those determined by \cite{Stverak}, who found $(n_{s}+n_{h})/n_{c}$ = $\sim$0.1 and 0.04-0.05 in fast and slow wind respectively. This test confirms that our algorithm is capable of differentiating between solar wind electron populations to a similar degree as previous results, with a very different method.

\section{Statistical Study}
\subsection{Methodology}

We then used ten years of pristine solar wind data, from 2001 to 2010, to quantify the relationship between strahl and halo breakpoint energies and other solar wind parameters, notably solar wind speed and core temperature. By quantifying the halo and strahl breakpoint energies separately, we determine if each suprathermal population is governed to the same extent by ambient conditions, or if they scale with each bulk parameter differently. For this study, we use Cluster-PEACE data in units of phase space density and split the data into four-minute intervals. The average solar wind speed during each interval is recorded using CIS measurements.

To confirm that Cluster is in the pristine solar wind, we used Cluster-FGM measurements and a model of the Earth's bow shock position \citep{CHAO_bow}. We use this model to identify when the spacecraft is outside the bow shock and not magnetically connected to it. We ensure Cluster is magnetically disconnected from the Earth's bow shock by discarding times when the magnetic field vector at Cluster intersects with the bow shock surface at any point.

We calculate the halo breakpoint energy, during each four-minute interval, by applying K-means clustering to phase space density values at 90$^{\circ}$ pitch angles, over a range of energies from 19 eV to 240 eV. Calculating the strahl/core breakpoint energy entails applying these K-means models to pitch angles and intervals which contain strahl. We achieve this by classifying flux-energy distributions during each interval, using the method in Section \ref{sec:Separating Halo and Strahl}, to determine if strahl is present at 0$^{\circ}$ or 180$^{\circ}$. 

We fit a Maxwellian velocity distribution function \citep{stverak_2008} to core velocities below each strahl or halo breakpoint energy, to determine the core temperature at that particular pitch angle. This function takes the form:
\begin{equation}
f_{c}=n_{c}\left (  \frac{m}{2\pi k} \right )^{3/2}\frac{1}{T_{c\perp}\sqrt{T_{c\parallel}}}\exp{\left [  -\frac{m}{2k}\left ( \frac{v_{\perp}^{2}}{T_{c\perp}}+\frac{v_{\parallel}^{2}}{T_{c\parallel}} \right ) \right ]},
\label{eq:core_vdf}
\end{equation}
where $n_{c}$ is the core density, $m$ the electron mass, $k$ is Boltzmann's constant, $T_{c\perp}$ and $T_{c\parallel}$ are the core perpendicular and parallel temperatures and $v_{\perp}$ and $v_{\parallel}$ are the perpendicular and parallel velocities. 

\subsection{Results}

\begin{figure}[ht]
\centering
\includegraphics[width=\hsize]{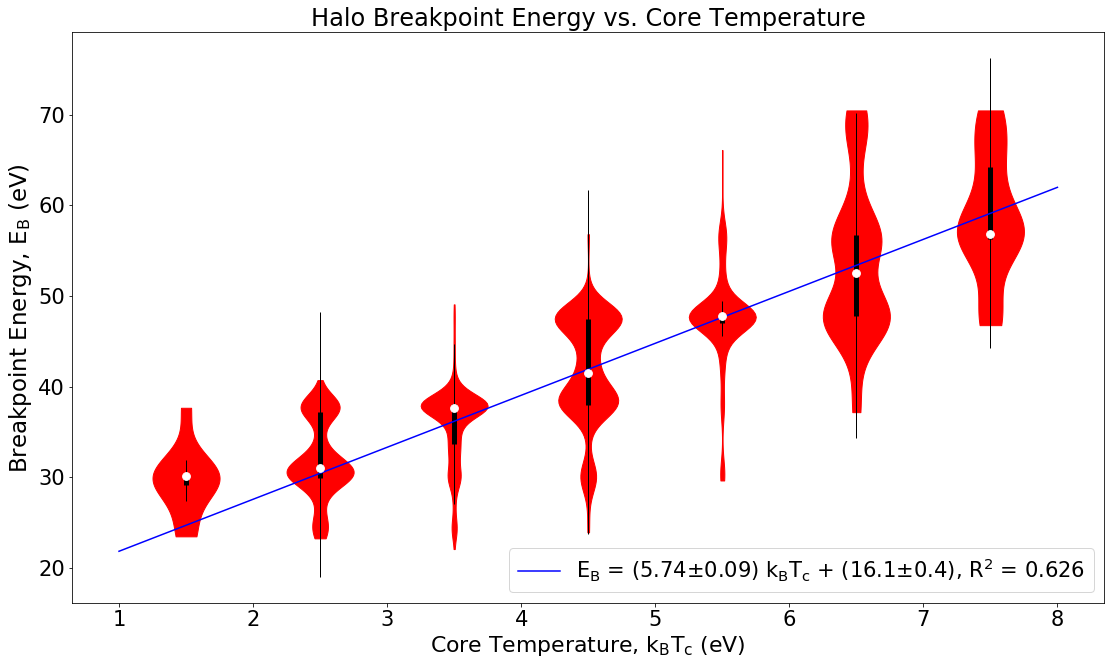}
\caption{`Violin plot' of halo breakpoint energy against core temperature. The blue line shows the line of best fit. The white dots indicate the median of breakpoint energies and the thick black lines show the inter-quartile ranges (IQR). We plot the thin black lines to display which breakpoint energies are outliers. They span from Q3+1.5$\times$IQR to Q1-1.5$\times$IQR, where Q3 and Q1 are the upper and lower quartiles, respectively. The horizontal width of the red regions represents the density of data points at that given breakpoint energy.}
\label{fig:halo_bp_tc_violin}
\end{figure}

Figure \ref{fig:halo_bp_tc_violin} shows the halo breakpoint energy vs. core temperature distribution in a `violin plot' to visualise the distribution of data points after binning the data into widths of 50 km/s. A violin plot is similar to a box plot, with the addition that the horizontal extend of each violin element represents a density plot of the data at different values. The red regions in Figure \ref{fig:halo_bp_tc_violin} visualise these density plots.

\begin{figure}[ht]
\centering
\includegraphics[width=\hsize]{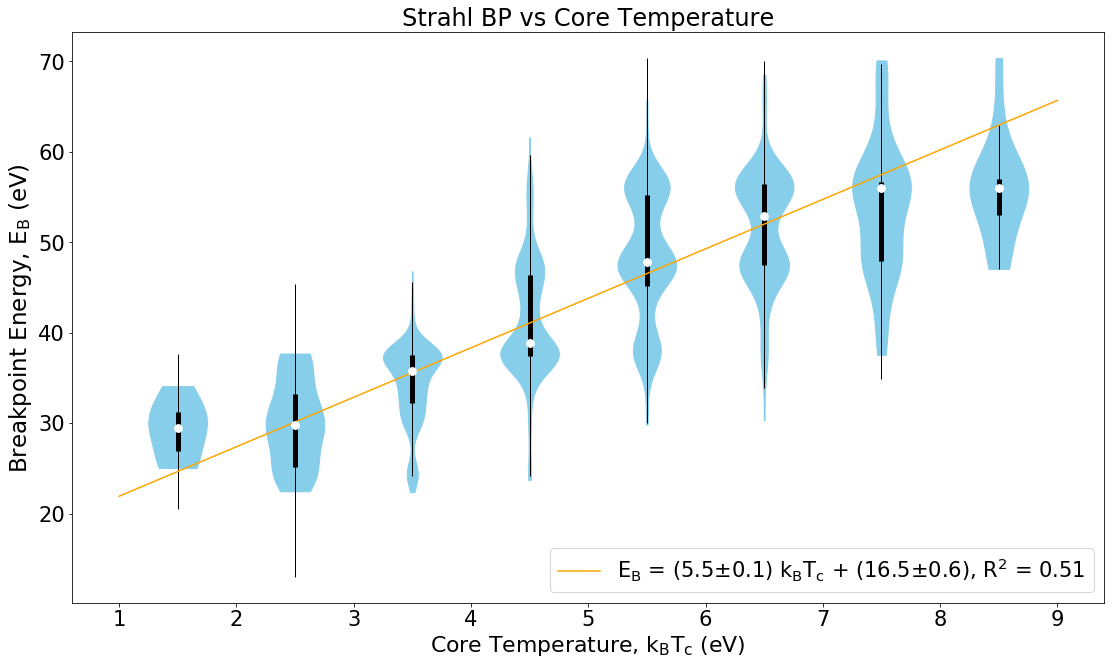}
\caption{`Violin plot' of strahl breakpoint energy against core temperature. The orange line shows the line of best fit. The remaining features are the same as in Figure \ref{fig:halo_bp_tc_violin}.}
\label{fig:strahl_bp_tc_violin}
\end{figure}

The widths of the red regions show that data are clustered about certain energies across all wind speeds. These regions of higher density in fact point to the energy channels (30.1 eV, 37.7 eV, 47.9 eV, 56.7 eV and 70.5 eV) within the C2-PEACE instrument's dataset. Figure \ref{fig:halo_bp_tc_violin} shows a clear positive correlation between halo breakpoint energy and core temperature, k\textsubscript{B}T\textsubscript{c}, with a gradient of 5.74$\pm$0.09. A statistical P-test produces a p-value of $<$0.0001, showing this relationship is significant at the p = 0.05 (5\%) level \citep{p-value}. The R-squared value of 0.626 indicates $\sim$63\% of variation in halo breakpoint energy can be described by this correlation. Very small inter-quartile ranges are observed in the 1-2 eV and 5-6 eV bins, while large inter-quartile ranges are observed in bins 4-5 eV and 6-7 eV. The results for the strahl breakpoint energy vs. core temperature are shown in Figure \ref{fig:strahl_bp_tc_violin}.

In both Figures \ref{fig:halo_bp_tc_violin} and \ref{fig:strahl_bp_tc_violin}, there is small discrepancy between the line of best fit and the median at core temperatures between 2 eV and 8 eV. When T\textsubscript{c} $<$ 2 eV, the linear fit underestimates all of the measured breakpoint energies, lying below the lower quartile range in both cases. In the strahl's case, the median and upper quartile at T\textsubscript{c} $>$ 8 eV drop significantly below the line of best fit. Figure \ref{fig:strahl_bp_tc_violin} suggests the dependence between core temperature and halo and strahl breakpoint energies differs. This is evidenced by the strahl breakpoint energy relation exhibiting a smaller gradient (5.5$\pm$0.1) and larger variance, based on the R-squared value of 0.51, with T\textsubscript{c} than the halo's relation. A p-value of $<$0.0001 suggests that this positive correlation between strahl breakpoint energy and core temperature is also highly significant at the p = 0.05 level.

\begin{figure}[ht]
\centering
\includegraphics[width=\hsize]{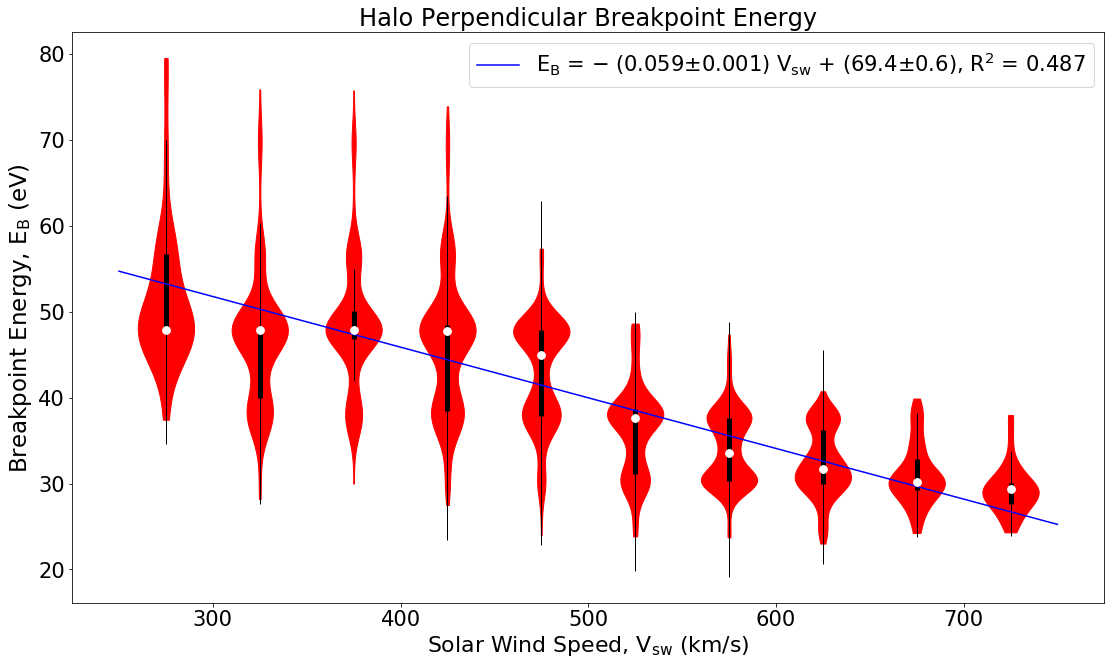}
\caption{`Violin plot' of halo breakpoint energy against solar wind speed. The blue line shows the line of best fit. The remaining features are the same as in Figure \ref{fig:halo_bp_tc_violin}.}
\label{fig:halo_perp_psd_violin}
\end{figure}

Figure \ref{fig:halo_perp_psd_violin} shows the results of our study to determine the relationship between halo breakpoint energy and solar wind speed. The collisionality of the solar wind plasma varies with its velocity, with slow wind typically exhibiting a higher collisionality than fast wind \citep{Scudder,Lie-Svendsen,Salem_2003,Gurgiolo}. Therefore, comparing breakpoint energy to solar wind velocity provides useful information on the scaling of breakpoint energy with the collisionality of the ambient plasma. Solar wind velocity is also a good indicator of the origin of the solar wind \citep{Geiss_95,Habbal_1997}, enabling us to investigate if breakpoint energy profiles vary with differing solar wind source regions. The gradient in Figure \ref{fig:halo_perp_psd_violin} is -5.9$\pm$0.1 eV per 100 km/s. The R-squared value of 0.487 is lower than 0.626 in Figure \ref{fig:halo_bp_tc_violin}, indicating that halo breakpoint energy exhibits a stronger correlation with core temperature than with solar wind speed. A statistical P-test produces a p-value of $<$0.0001, showing this relationship is significant at the p = 0.05 (5\%) level.

The distribution of breakpoint energies with wind speed in Figure \ref{fig:halo_perp_psd_violin} displays a step function at about 500 km/s. The lower quartile within the 450-500 km/s bin lies above the upper quartiles in faster speed bins. Fitting two linear fits to solar wind speeds below and above 500 km/s separately produces gradients of -4.2$\pm$0.1 eV per 100 km/s and -3.5$\pm$0.1 eV per 100 km/s respectively. The associated R-squared values are 0.588 and 0.651 respectively; both larger than a value 0.487 for a single linear fit, indicating that two separate correlations better describe the distribution in Figure \ref{fig:halo_perp_psd_violin} than a single correlation. The two correlations are also significant at the p = 0.05 (5\%) level. The data-points in Figure \ref{fig:halo_perp_psd_violin} are distributed along a larger range of breakpoint energies at lower wind speeds than higher wind speeds. However, according to the inter-quartile ranges for the majority of data-points, the variance about the median values is relatively small, with the exception of a few outliers. The medians themselves do not deviate significantly from the line of best fit across all wind speeds, with the largest median residual equalling 5 eV in the $<$300 km/s bin. There is some evidence for positive or negative skewness at certain solar wind velocities, such as in the $<$300 km/s and 400-450 km/s bins, as can be seen when the median appears to lie on one of the edges of the inter-quartile range. 

\begin{figure}[ht]
\centering
\includegraphics[width=\hsize]{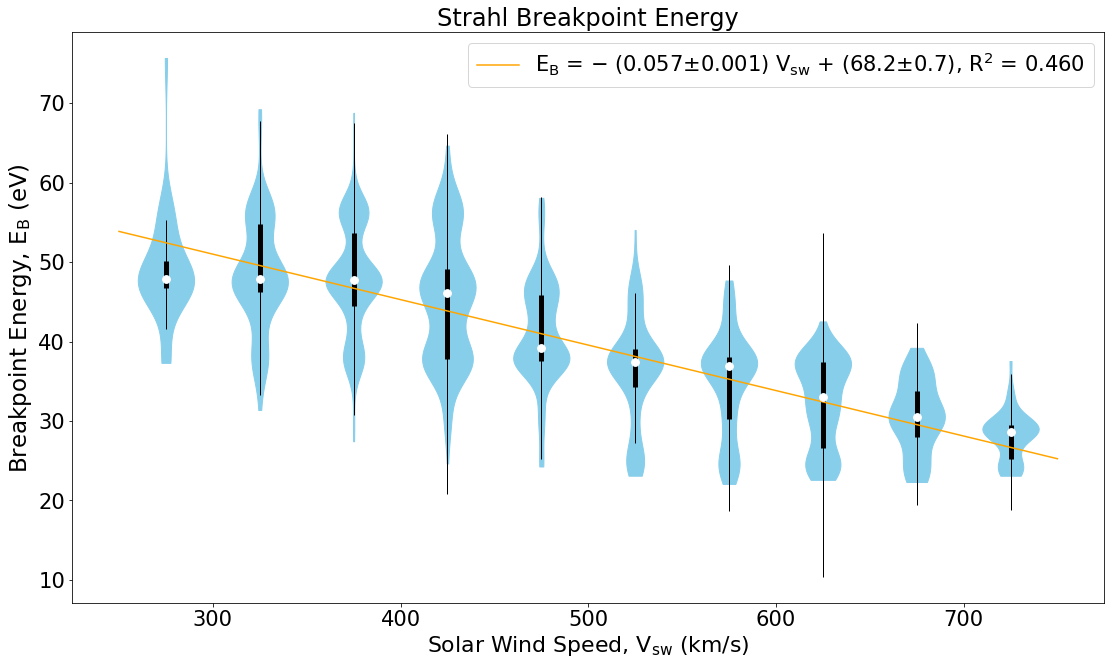}
\caption{`Violin plot' of strahl breakpoint energy against solar wind speed. The orange line shows the line of best fit. The remaining features are in the same format as Figure \ref{fig:halo_bp_tc_violin}.}
\label{fig:strahl_psd_violin}
\end{figure}

Figure \ref{fig:strahl_psd_violin} shows the strahl breakpoint energy variation with solar wind speed. According to our linear fit, the rate of decrease of strahl breakpoint energy with solar wind speed is -5.7$\pm$0.1 eV per 100 km/s. Solar wind speed has a smaller correlation with strahl breakpoint energy than halo breakpoint energy, based on the steepness of each gradient and R-squared values. This R-squared value of 0.460 in Figure \ref{fig:strahl_psd_violin} also indicates that the strahl breakpoint energy has a weaker correlation statistically with solar wind speed than with core temperature, as the line of best fit describes less of the variation. This is also the case for the halo breakpoint energy. A p-value of $<$0.0001 indicates that this negative correlation is also highly significant at the p = 0.05 level.

Similar to Figure \ref{fig:halo_perp_psd_violin}, the variation in breakpoint energy in the strahl violin plot is larger at smaller wind speeds. However, unlike for halo, the 400-450 km/s bin has a much larger variance than the $<$300 km/s bin, as evidenced by their inter-quartile ranges. This larger spread of data at medium wind speeds explains why the strahl's R-squared value is lower than the halo's. The lack of skewness in Figure \ref{fig:strahl_psd_violin} shows that the data are distributed more symmetrically in the strahl's case than the halo's. The sum of the median residuals are also smaller for the strahl, with the largest median residual at 3.5 eV in the $<$300 km/s solar wind speed bin. A step function is less apparent in Figure \ref{fig:strahl_psd_violin}, however there is a clear distinction between the median breakpoint energy relation with wind speed in slow winds ($<$450 km/s), compared to fast winds. Table \ref{tab:gradients_correlations} contains the gradients and R-squared values of the correlations in Figures \ref{fig:halo_bp_tc_violin}, \ref{fig:strahl_bp_tc_violin}, \ref{fig:halo_perp_psd_violin}, and \ref{fig:strahl_psd_violin}.

\begin{table}[ht]
\caption{Correlations between halo and strahl breakpoint energies with core temperature, $T_{c}$ and solar wind speed, $V_{sw}$, as represented by the gradients and R-squared, $R^{2}$, values.}
\centering
\begin{tabular}{l|cc|cc}
                            & \multicolumn{2}{c|}{$T_{c}$} & \multicolumn{2}{c}{$V_{sw}$} \\ \cline{2-5} 
\multicolumn{1}{c|}{Population}   & $E_{bp}/T{c}$        & $R^{2}$        & $E_{bp}/V_{sw}$         & $R^{2}$        \\ 
\multicolumn{1}{c|}{} & & & [eV/(km/s)] & \\ \hline
\multicolumn{1}{c|}{Halo}   & 5.74         & 0.626       & -0.059           & 0.487       \\ \hline
\multicolumn{1}{c|}{Strahl} & 5.5         & 0.51       & -0.057           & 0.460      
\end{tabular}
\label{tab:gradients_correlations}
\end{table}

\section{Discussion}

In this study, we use the K-means algorithm to successfully distinguish between the three populations and we train a supervised learning algorithm (K-nearest neighbours) to classify a subset of the pitch angle and energy distributions. There is a strong agreement between the two machine learning methods, allowing us to apply the K-means clustering method to a larger subset of solar wind electron data at different solar wind velocities. Machine learning algorithms provide us with an efficient method of classification from which small scale variations of electron populations in relation to energy and pitch angle can be derived. By classifying a single distribution at each time step, we build up a high resolution picture of suprathermal breakpoint energy and relative number density, including how they evolve with different parameters. The techniques we employ can be easily applied to any classification problem where sufficient data are available.

Distinguishing between strahl, halo, and core electron populations allows us to calculate their relative number densities, in order to compare our method to previous results. \cite{Stverak} show that suprathermal electrons in the fast wind constitute $\sim$10\% of the total electron number density, while in  slow wind they occupy 4\% to 5\% of the total electron density. In comparison, we obtain values of $\sim$9.4\% and 2.5-4.3\% for fast and slow wind respectively. Obtaining densities of the same order as \cite{Stverak} confirms that our method is capable of distinguishing between multiple solar wind electron populations to a similar degree as alternative methods. Being a zeroth order moment, there is a smaller level of uncertainty when calculating the density, as opposed to the breakpoint energy or higher order moments, by fitting distribution functions. Using machine learning techniques instead of fitting bi-Maxwellian and bi-Kappa functions to electron velocity distributions, which involves fixing certain parameters \citep{Stverak}, eliminates the need to use prior assumptions about these solar wind electron populations. Therefore, our new method results in more robust estimations of the solar wind electrons’ breakpoint energies.

The observation that the majority of the halo population is formed due to strahl scattering \citep{saito_gary,pagel,Stverak} explains the relationship between $n_{s}/n_{h}$ and wind speed in Figure \ref{fig:wind_speed_ratios_labels}. Strahl in slow solar wind undergoes more scattering per unit distance than in faster wind \citep[e.g.][]{Fitzenreiter}, leading to a higher value of $n_{h}/n_{e}$ at 1 au. We observe a near absence of strahl in very slow solar wind at velocities of 308 km/s (see Figures \ref{fig:3d_classification_slow_wind} and \ref{fig:wind_speed_ratios_labels}), which is consistent with observations from previous studies \citep[e.g.][]{Fitzenreiter,Gurgiolo,Graham_2018}. By analysing a number of periods of slow solar wind, \cite{Fitzenreiter} find that the strahl generally has a larger width in slow solar winds than fast, while \cite{Gurgiolo} find that strahl is often not present at solar wind velocities $\lesssim$ 425 km/s. \cite{Graham_2018} also note an absence of strahl during certain slow solar wind times. This absence of strahl remains unexplained. Possible hypotheses include: Coulomb pitch angle scattering which counteracts magnetic focussing effects during strahl formation \citep{horaites}, intense scattering due to broadband whistler turbulence \citep{Pierrard2001}, and the lack of initial strahl formation during the production of slow solar wind \citep{Gurgiolo}.

Instead of finding the intersection between core and suprathermal fitting functions \citep[e.g.][]{Pilipp_evdf,mccomas_92,Stverak}, a method which according to \cite{mccomas_92} produces `somewhat arbitrary' values, our method calculates the breakpoint energy based on the data recorded in each individual pitch angle and energy bin. Our method calculates breakpoint energy values of both sunward and anti-sunward strahl, occasionally obtaining two strahl breakpoint energy values at a single time if bi-directional strahl is present. An alternative method is presented by \cite{Stverak} who discard sunward strahl in their calculations of the strahl $E_{bp}$/$k_{B}T_{c}$ ratio at each radial distance. By characterising both sunward and anti-sunward strahl, our method significantly improves the characterisation of all electron beams in the solar wind. 

Our work on the core velocity distribution functions elucidates the relative correlation between core temperature, T\textsubscript{c}, and both halo and strahl breakpoint energies. Using core temperature as a reference point enables us to predict to what extent strahl and halo characteristics scale to characteristics of the core. The core temperature has a strong correlation with both suprathermal breakpoint energies, with the halo breakpoint energy exhibiting a closer correlation than the strahl's. Both halo and strahl breakpoint energies statistically have a stronger correlation with core temperature than with solar wind speed. The gradients between breakpoint energy and core temperature are calculated as 5.74$\pm$0.09 and 5.5$\pm$0.1 for halo and strahl respectively. 

The linear relationship that we observe between breakpoint energy and core temperature is in line with previous measurements \citep[e.g.][]{mccomas_92,Stverak}, for both the halo and strahl. According to \cite{Scudder}, a linear trend in the halo relation also follows under the assumption that binary Coulomb collisions dominate electron dynamics in the solar wind. However, in order to align with available experimental data, \cite{Scudder} set a scaling factor of $E_{bp}/k_{B}T_{c}$ = 7, which differs from our scaling factor of $E_{bp}/k_{B}T_{c}$ = 5.5$\pm$0.1. With a scaling factor of $E_{bp}/k_{B}T_{c}$ = 7, \cite{Scudder} predict that a transformation of thermal electrons into the suprathermal population occurs as the solar wind flows out from the Sun. Findings by \cite{Stverak}, on the other hand, show that the $(n_{h}+n_{s})/n_{c}$ ratio remains roughly constant with heliocentric distance in the slow wind, suggesting a lack of interchange between the thermal and suprathermal populations. However \cite{Stverak} observes some variability in the $(n_{h}+n_{s})/n_{c}$ ratio in the fast wind, which they attribute to either statistical effects due to a lack of samples or a possible `interplay' between thermal and suprathermal electrons. \cite{Scudder} also predict that the halo $E_{bp}/k_{B}T_{c}$ ratio remains constant with heliocentric distance, whereas \cite{Stverak} find that the halo $E_{bp}/k_{B}T_{c}$ ratio decreases with heliocentric distance. These findings by \cite{Stverak}, along with the discrepancy between our calculated ratio of $E_{bp}/k_{B}T_{c}$ = 5.5$\pm$0.1 and the prediction of $E_{bp}/k_{B}T_{c}$ = 7, suggest that the model of \cite{Scudder} requires a minor update to either the theory or to the input parameters. The discrepancy, however, may also be indicative of other processes, such as wave-particle scattering \citep[e.g.][]{gary_94}, that possibly modifies the ratio between breakpoint energy and core temperature while preserving its linear relationship.

In our statistical study, we find that both strahl and halo breakpoint energies decrease with solar wind speed. At all solar wind velocities, as well as core temperatures, the halo breakpoint energy is larger than the strahl's at equivalent velocities and temperatures. The halo breakpoint energy exhibits a higher correlation with the solar wind speed than strahl. The anti-correlation between the two parameters corresponds with the finding that $(n_{h}+n_{s})/n_{c}$ increases with solar wind speed \citep{Stverak}, where $n_{h}$, $n_{s}$ and $n_{c}$ represent the halo, strahl, and core number densities. Assuming all plasma parameters are kept constant, except for the core density and temperature, the relative density of suprathermal electrons will increase if the breakpoint energy decreases. This observed relationship between solar wind speed and electron ratios is most likely a result of the lower collisionality of fast solar wind \citep{Scudder,Lie-Svendsen,Salem_2003,Gurgiolo}, which results in more distinctive non-thermal features of the electron velocity distribution function. Further work is required to analyse whether different breakpoint energy relations exist that depend on the source of solar wind. Initial findings in this paper suggest the existence of two distinct relationships in the halo breakpoint energy vs. wind speed distribution, with a step function at 500 km/s. This finding links to a sharp distinction between fast and slow solar winds \citep{feldman_origin}. Therefore the origin of the solar wind, i.e., coronal holes for fast wind or streamer belt regions for slow wind, potentially plays a role in the definition of thermal and non-thermal electron populations. A step function is less obvious in the strahl breakpoint energy vs. solar wind speed distribution.

\section{Conclusions}

In this study, we apply unsupervised K-means clustering algorithms to Cluster-PEACE data to separate solar wind electron pitch angle and energy distributions into the core, halo, and strahl populations. This enables us to perform an accurate statistical analysis of strahl and halo breakpoint energies. In our statistical study, we compare the relationship between core temperature, $T_{c}$ and both halo and strahl breakpoint energies. We present a strong correlation between suprathermal breakpoint energies and $T_{c}$, and conclude this is due to core temperature being a determining factor for breakpoint energy. As a result of higher core temperatures, the Maxwellian part of the total electron velocity distribution function, which represents the core, extends across a wider range of velocity space \citep{Pilipp_evdf}. The core distribution therefore overlaps with the halo and strahl at higher energies and thus increases the suprathermal breakpoint energy. 

We find that halo breakpoint energy remains larger than the strahl’s across all temperatures. This difference between halo and strahl breakpoint energies suggests that there are certain energies, below the halo breakpoint energy, at which a strahl and core population are both present. At these energies, strahl dominates at parallel pitch angles and core dominates at perpendicular pitch angles. Wave-particle scattering processes \citep{gary_94,Vasko_2019,Verscharen_2019} scatter these low energy strahl electrons to higher perpendicular velocities and smaller parallel velocities. At sufficiently high core temperatures, these strahl electrons would be absorbed by the core population \citep{Pilipp}, instead of the higher energy halo population. The absorption of strahl electrons by the core increases the number of Coulomb collisions \citep{Landi_2012}, which then leads to an increase in core temperature \citep{Marsch,Boldyrev}. This scenario is consistent with previous studies \citep{pilipp_anisotropy} which show a transfer of electron kinetic energy from the parallel to perpendicular direction, increasing core temperature in the perpendicular direction. The increase of core temperature, due to the absorption of strahl electrons, acts to extend the core component of the electron velocity distribution function to higher velocities \citep{Pilipp_evdf}, therefore increasing the halo breakpoint energy at pitch angles at which the strahl is not present. This phenomenon explains the larger difference between strahl and halo breakpoint energies at higher core temperatures, as a larger difference in breakpoint energy means more strahl electrons are scattering into the core population rather than the halo population.

This work signifies the first extensive study in characterising the relation between breakpoint energy and solar wind speed, for each of the suprathermal populations. Our results show there is a significant decrease in both halo and strahl breakpoint energies with increasing solar wind speed, with the halo relation exhibiting a stronger correlation. We find two distinct relationships in the halo breakpoint energy vs. solar wind speed distribution, with a step function at 500 km/s. We predict this step function relates to the difference in origin of fast and slow solar wind electrons \citep{feldman_origin}. Further investigation, with the aid of new facilities provided by the Parker Solar Probe and Solar Orbiter missions, can test this prediction and investigate why the step function is prevalent in the halo breakpoint energy relationship but not in the strahl breakpoint energy relationship with solar wind speed. In future studies, using Solar Orbiter measurements at smaller heliocentric distances will allow us to better characterise halo and strahl breakpoint energies and improve our understanding of their dependence on bulk solar wind parameters.

\begin{acknowledgements}
M.R.B. is supported by a UCL Impact Studentship, joint funded by the ESA NPI programme. I.J.R., D.V. and A.W.S. are supported by STFC Consolidated Grant ST/S00240/1. D.V. is supported by the STFC Ernest Rutherford Fellowship ST/P003826/1. A.W.S. is supported by NERC grant NE/P017150/1. T.B. is supported by STFC Training Grant ST/R505031/1. C.E.J.W. is supported by STFC Grant ST/R000921/1 and NERC Grant NE/P017274/1. We thank the Cluster instrument teams (PEACE, FGM, CIS, EFW) for the data used in this study, in particular the PEACE operations team at the Mullard Space Science Laboratory. Data for the Cluster spacecraft can be obtained from the Cluster Science Archive (https://csa.esac.esa.int/csa-web/). We thank the anonymous reviewer for their many useful contributions to this manuscript. This work was discussed at the 2019 ESAC Solar Wind Electron Workshop, which was supported by the Faculty of the European Space Astronomy Centre (ESAC).
\end{acknowledgements}

\bibliography{A&A_Ref}

\end{document}